\documentclass{jpsj2}

\usepackage{bm}
\usepackage{braket}
\newcommand{\diff}{\mathrm{d}}
\newcommand{\subc}{_\mathrm{c}}
\newcommand{\subs}{_\mathrm{s}}

\newcommand{\pdif}[2]{\frac{\partial #1}{\partial #2}}

\newcommand{\pdifdif}[3]{\frac{\partial^2 #1}{\partial #2 \partial #3}}
\newcommand{\trp}[1]{{}^t #1}

\title{Neural network model with discrete and continuous information representation}

\author{Jun \textsc{Kitazono}$^{1}$\thanks{E-mail: kitazono@mns.k.u-tokyo.ac.jp}, 
        Toshiaki \textsc{Omori}$^{1,2}$, and 
        Masato \textsc{Okada}$^{1,2}$
        }

\inst{$^{1}$Graduate School of Frontier Science, The University of Tokyo\hskip1em
  5--1--5 Kashiwanoha, Kashiwa-shi, Chiba-ken 277--8561 \\
$^{2}$RIKEN Brain Science Institute\hskip1em
  2--1 Hirosawa, Wako-shi, Saitama-ken 351--0198 \\}

\abst{
An associative memory model and a neural network model with a Mexican-hat type interaction 
are the two most typical attractor networks used in the artificial neural network models. 
The associative memory model has discretely distributed fixed-point attractors, 
and achieves a discrete information representation. 
On the other hand, a neural network model with a Mexican-hat type interaction 
uses a line attractor to achieves a continuous information representation, 
which can be seen in the working memory in the prefrontal cortex and columnar activity in the visual cortex. 
In the present study, we propose a neural network model 
that achieves discrete and continuous information representation. 
We use a statistical-mechanical analysis to find that a localized retrieval phase exists in the proposed model, 
where the memory pattern is retrieved in the localized subpopulation of the network. 
In the localized retrieval phase, the discrete and continuous information representation is achieved 
by using the orthogonality of the memory patterns and 
the neutral stability of fixed points along the positions of the localized retrieval. 
The obtained phase diagram suggests that 
the antiferromagnetic interaction and the external field are important for 
generating the localized retrieval phase. 
}

\kword{Mexican-hat type interaction, Associative memory model, Localized activity, Statistical mechanics}

\begin{document}
\maketitle

\section{Introduction}
An associative memory model and a neural network model with a Mexican-hat type interaction 
are the two most typical attractor networks used in artificial neural network models. 
The associative memory model represented by the Hopfield model has discretely distributed fixed-point attractors \cite{Hopfield}. 
On the other hand, the model with a Mexican-hat type interaction, 
modeled after the hyper-column of the primary visual cortex and the frontal cortex during the memory-guided saccade, 
has continuously distributed fixed-point attractors (what is termed a line attractor)
that reflect the distant relationship between inputs\cite{Ben-Yishai, Camperi, Hamaguchi, Urano}. 

Recent electrophysiological studies suggest that 
neurons represent informations as sparse and local neuronal excitations 
in the inferior temporal cortex (IT) \cite{Wang, Tamura}, 
specifically, neurons in IT area achieve a discrete and continuous information representation 
by using both positions of a local excitation and microscopic firing patterns of a sparse activity in the local excitation. 
For instance, neurons measure angles of a visual stimulus with the positions of a local excitation, and consequently 
the corresponding information representation (firing pattern) continuously changes 
when the visual stimulus is rotated \cite{Wang}.
Meanwhile, 
neurons discriminate visual stimuli by differences between microscopic firing patterns of a sparse activity, and consequently 
the information representation discretely changes 
when the visual stimulus is replaced \cite{Tamura}. 
Based on these evidence, Wada \textit{et al.}\cite{Wada} proposed a self-organizing map (SOM) model of IT area, and 
confirmed by numerical experiment that the sparse and local neuronal excitation is self-organized in their model. 
Additionally, Hamaguchi \textit{et al.}\cite{Hamaguchi, Urano} proposed an Ising spin neural network model 
with a disordered Mexican-hat type interaction, and 
verified that the sparse and local neuronal excitation is achieved in their model. 
However, these models do not store more than one pattern. 
In addition, Ichiki \textit{et al.}\cite{ˆê–Ø} analyzed a model with spatially modulated Hebbian interactions. 
Equilibria of their model are transformed into those of the model with the Mexican-hat type interaction\cite{Hamaguchi, Urano} 
by a gauge transformation based on a stored pattern. 

In this study, we propose a solvable neural network model, 
that achieves the discrete and continuous information representation. 
Our model is based on a Hebbian interaction weighted by a Mexican-hat type interaction \cite{ˆê–Ø}. 
We use a statistical-mechanical analysis to show that a localized retrieval (LR) phase exists in the proposed model, 
where a localized part of the stored pattern is retrieved. 
The LR states are discretely and continuously distributed in the network state space, 
that is, our model has discretely distributed line attractors. 
Thus, the LR states correspond to the discrete and continuous information representation. 
Additionally, we use a stability analysis and a phase diagram to 
we find that the stability of the LR states depends on the number of the stored patterns, and 
that an antiferromagnetic interaction and a negative external magnetic field are essential for stabilizing the LR states. 

\section{Model}
\begin{figure}
\includegraphics[width=3.2in]{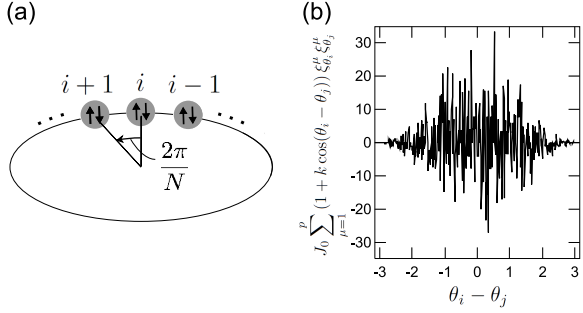}
\caption{
Schematic of our model. 
(a) Geometry of neurons. The $N$-neurons are circularly distributed at regular intervals. 
(b) Schematic of one part of interneuronal interaction 
$J_0\sum_{\mu=1}^p (1+k\cos(\theta_i-\theta_j))\xi_{\theta_i}^{\mu}\xi_{\theta_j}^{\mu}$ 
(the first part of the right side of Eq. (\ref{eq:Jij})) 
as a function of neuronal interval $\theta_i-\theta_j$. 
This part represents a Hebbian interaction weighted by a Mexican-hat type interaction. 
$J_0$ represents the strength of the Hebbian interaction, and $k$ represents the strength of the weighting 
by the Mexican-hat type interaction. 
}
\label{fig:model}
\end{figure}

We analyze an Ising spin neural network model with a microscopic state defined by the $N$-neuron state vector 
$\bm{S}=(S_{\theta_1},S_{\theta_2},\cdots,S_{\theta_N})\in\{-1,1\}^N$. 
Here $S_{\theta_i}=1$ if neuron $i$ fires, and $S_{\theta_i}=-1$ if it is at rest. 
Neuron $i$ is located at angle $\theta_i = \frac{2\pi i}{N}-\pi$ on a one-dimensional ring 
indexed by $\theta_i \in (-\pi,\pi]$, as shown in Fig. \ref{fig:model}(a). The Hamiltonian of the system 
we are going to study is
\begin{equation}
 H(\bm{S})=-\frac{1}{2}\sum_{i, j}J_{\theta_i\theta_j}S_{\theta_i} S_{\theta_j} - h \sum_{i} S_{\theta_i} ,\label{eq:Hamiltonian}
\end{equation}
where $h$ is an external magnetic field (representing a common external input to neurons). 
The interaction $J_{\theta_i\theta_j}$ (representing the synaptic interaction between the $i$th neuron and the $j$th neuron)
consists of a Hebbian interaction weighted by a Mexican-hat type interaction, and 
the antiferromagnetic interaction:
\begin{equation}
 J_{\theta_i\theta_j}
 =\frac{J_0}{N}\sum_{\mu=1}^{p}(1+k\cos(\theta_i-\theta_j))\xi_{\theta_i}^{\mu}\xi_{\theta_j}^{\mu}
 -\frac{g}{N}, \label{eq:Jij}
\end{equation}
where $J_0$ represents the strength of the Hebbian interaction, $k$ represents the strength of the weighting 
by the Mexican-hat type interaction, and $g$ represents the strength of the antiferromagnetic interaction. 
$\xi_{\theta_i}^\mu$ denotes the $i$th component of the stored pattern vector $\bm{\xi}^\mu \in \{-1,1\}^N$, 
and is a quenched independent random variable taking $1$ or $-1$ with a probability $\frac{1}{2}$. 
$p$ is the number of stored patterns. 
We restrict ourselves to a case where $p$ is finite in the thermodynamic limit $N\to\infty$ ($p=O(1)$) throughout this paper. 
The Hopfield model in our model\cite{Hopfield} is one where $k=g=h=0$. 
The model with the Mexican-hat type interaction in our model\cite{Ben-Yishai, Hamaguchi, Urano} is one where 
$k\to\infty$ with $J_0 k$ remaining finite, $g=0$, $p=1$, and $\xi_{\theta_i}^1=1(1\leq \forall i \leq N)$. 
Thus, our model includes the Hopfield model and the model with the Mexican-hat type interaction as special cases. 
In addition, the case where $J_0<0,k<0$ and $g=h=0$ in our model has been previously studied \cite{ˆê–Ø}. 
The antiferromagnetic interaction $g$ and the external magnetic field $h$ play significant roles in generating 
the localized retrieval phase, where a localized part of a stored pattern is retrieved, 
as described later. 

\section{Statistical-Mechanical Analysis}
\subsection{Free Energy and Hessian Matrix}
In this section, we calculate the free energy per neuron and the Hessian matrix. 
We define the following order parameters as: 
\begin{align}
  m(\bm{S})&=\frac{1}{N}\sum_i S_{\theta_i}, \label{eq:m}\\
  m_0^\mu(\bm{S})&=\frac{1}{N}\sum_i S_{\theta_i} \xi_{\theta_i}^\mu, \label{eq:m0}\\
  m\subc^\mu(\bm{S})&=\frac{1}{N}\sum_i S_{\theta_i}\xi_{\theta_i}^\mu \cos\theta_i, \label{eq:mc}\\
  m\subs^\mu(\bm{S})&=\frac{1}{N}\sum_i S_{\theta_i}\xi_{\theta_i}^\mu \sin\theta_i, \label{eq:ms}
\end{align}
$m_0^\mu$ is the overlap between the spin configuration and the stored pattern, and 
$m\subc^\mu$ and $m\subs^\mu$ are the order parameters that are 
the first Fourier coefficients of $\{S_{\theta_i} \xi_{\theta_i}^\mu\}_{i=1}^N$, regarded as a function of $\theta$.
Additionally we introduce another two order parameters for later use:
\begin{align}
  m_1^\mu(\bm{S}) &= \sqrt{m\subc^\mu(\bm{S})^2 + m\subs^\mu(\bm{S})^2}, \\
  \phi^\mu(\bm{S}) &=\arctan(m\subs^\mu(\bm{S}) /m\subc^\mu(\bm{S})),
 \end{align}
where $m_1^\mu$ and $\phi^\mu$ are the amplitude and phase of the first Fourier component of 
$\{S_{\theta_i} \xi_{\theta_i}^\mu\}_{i=1}^N$, respectively. 
The Hamiltonian Eq. (\ref{eq:Hamiltonian}) is represented by these order parameters as: 
\begin{equation}
 \begin{split}
 &H\bigl(m(\bm{S}),\{m_0^\mu(\bm{S})\},\{m\subc^\mu(\bm{S})\},\{m\subs^\mu(\bm{S})\}\bigr) \\
 =&N\left(-\frac{J_0}{2}\sum_{\mu=1}^p \left\{ m_0^\mu(\bm{S})^2 + k \left(m\subc^\mu (\bm{S})^2 + m\subs^\mu(\bm{S})^2\right) \right\}
         +\frac{g}{2} m(\bm{S})^2 - hm(\bm{S}) \right) . 
 \end{split}
\end{equation}
The partition function $Z=\mathrm{Tr}\, \mathrm{e}^{-\beta H}$ is represented by the delta function as:
\begin{equation}
 \begin{split}
 Z =&\mathrm{Tr}_{\bm{S}} \int_{-\infty}^\infty \cdots\int_{-\infty}^\infty\diff m \prod_{\mu=1}^{p}\diff m_0^\mu \diff m\subc^\mu \diff m\subs^\mu \\
    &\delta\bigl(m-m(\bm{S})\bigr) \prod_{\mu=1}^{p} \delta\bigl(m_0^\mu-m_0^\mu(\bm{S})\bigr)\delta\bigl(m\subc^\mu-m\subc^\mu(\bm{S})\bigr)\delta\bigl(m\subs^\mu-m\subs^\mu(\bm{S})\bigr) \\
   &\exp\Bigl(-\beta H\bigl(m,\{m_0^\mu\},\{m\subc^\mu\},\{m\subs^\mu\}\bigr) \Bigr). \label{eq:partition}
 \end{split}
\end{equation}
We can rewrite Eq. (\ref{eq:partition}) by the Fourier integral representation of the delta function
\begin{equation}
 \delta\bigl(m^\prime-m^\prime(\bm{S})\bigr)=\frac{N}{2\pi}\int_{-\infty}^{\infty}\diff \hat{m}^\prime \exp(iN\hat{m}^\prime(m^\prime-m^\prime(\bm{S}))),
\end{equation}
substituting  $m$, 
              $m_0^\mu$, 
              $m\subc^\mu$, and 
              $m\subs^\mu$ 
              for $m^\prime$,
              and
              $\frac{1}{N}\sum_i S_{\theta_i}$, 
              $\frac{1}{N}\sum_i S_{\theta_i} \xi_{\theta_i}^\mu$, 
              $\frac{1}{N}\sum_i S_{\theta_i} \xi_{\theta_i}^\mu \cos\theta_i$, and 
              $\frac{1}{N}\sum_i S_{\theta_i} \xi_{\theta_i}^\mu \sin\theta_i$ 
              for $m^\prime(\bm{S})$,                
respectively;
\begin{equation}
 \begin{split}
  Z =&\mathrm{Tr}_{\bm{S}} \int_{-\infty}^\infty \cdots\int_{-\infty}^\infty\diff m \prod_{\mu=1}^{p}\diff m_0^\mu \diff m\subc^\mu \diff m\subs^\mu \diff \hat{m} \prod_{\mu=1}^{p}\diff \hat{m}_0^\mu \diff \hat{m}\subc^\mu \diff \hat{m}\subs^\mu\\
     &\exp\left[-\beta H\bigl(m,\{m_0^\mu\},\{m\subc^\mu\},\{m\subs^\mu\}\bigr)
               +iN\left(\sum_{\mu=1}^p \bigl(\hat{m}_0^\mu m_0^\mu + \hat{m}\subc^\mu m\subc^\mu + \hat{m}\subs^\mu m\subs^\mu \bigr) + \hat{m}m\right)\right.\\
         &\left.-i\sum_{i=1}^N S_{\theta_i} \left(\sum_{\mu=1}^p \xi_{\theta_i}^\mu (\hat{m}_0^\mu+\hat{m}\subc^\mu\cos\theta_i+\hat{m}\subs^\mu\sin\theta_i) + \hat{m}\right)
                \right] .\label{eq:ZbeforeTr}
 \end{split}
\end{equation}
Taking the trace of Eq. (\ref{eq:ZbeforeTr}), we get
\begin{equation}
 \begin{split}
  Z =&\int_{-\infty}^\infty \cdots\int_{-\infty}^\infty \diff m \prod_{\mu=1}^{p}\diff m_0^\mu \diff m\subc^\mu \diff m\subs^\mu \diff \hat{m} \prod_{\mu=1}^{p}\diff \hat{m}_0^\mu \diff \hat{m}\subc^\mu \diff \hat{m}\subs^\mu \\
     &\exp\left[-\beta H\bigl(m,\{m_0^\mu\},\{m\subc^\mu\},\{m\subs^\mu\}\bigr) \phantom{\left(\sum_{\mu=1}^p\right)} \right.\\
               & \left. +iN\left(\sum_{\mu=1}^p \bigl(\hat{m}_0^\mu m_0^\mu + \hat{m}\subc^\mu m\subc^\mu + \hat{m}\subs^\mu m\subs^\mu \bigr) + \hat{m}m\right) 
                +\sum_{i=1}^N \log 2\cosh\beta\tilde{h}(\theta_i)
              \right] .\label{eq:ZafterTr}
 \end{split}
\end{equation}
where  $\tilde{h}(\theta_i)$ is a local magnetic field defined as
\begin{equation}
  \tilde{h}(\theta_i) = -\frac{i}{\beta}\left(\sum_{\mu=1}^p \xi_{\theta_i}^\mu (\hat{m}_0^\mu+\hat{m}\subc^\mu\cos\theta_i+\hat{m}\subs^\mu\sin\theta_i) + \hat{m}\right). \label{eq:h_before}
\end{equation}
We can evaluate these integrals by using the steepest descent in the thermodynamic limit $N\to\infty$. 
\begin{equation}
  Z = \exp(-N\beta f),
\end{equation}
where $f$ is the free energy per neuron given as
\begin{equation}
  \begin{split}
    f =& -\frac{J_0}{2}\sum_{\mu=1}^p \left\{(m_0^\mu)^2 + k\left((m\subc^\mu)^2 + (m\subs^\mu)^2\right) \right\} +\frac{g}{2} m^2 - hm \\
       &-\frac{i}{\beta}\left(\sum_{\mu=1}^p \bigl(\hat{m}_0^\mu m_0^\mu + \hat{m}\subc^\mu m\subc^\mu + \hat{m}\subs^\mu m\subs^\mu \bigr) + \hat{m}m\right) -\frac{1}{2\pi\beta} \int_{-\pi}^{\pi} \diff \theta \log 2\cosh\beta\tilde{h}(\theta) .
    \label{eq:fe_before}   
  \end{split}
\end{equation}
The summation of $\theta_i$ is now replaced by an integral over $\theta$. 
The saddle point equations are 
\begin{align}
  \pdif{f}{m}         &=  gm - h - \frac{i}{\beta}\hat{m} = 0, \label{eq:saddle_m} \\
  \pdif{f}{m_0^\mu}   &= -J_0m_0^\mu - \frac{i}{\beta}\hat{m}_0^\mu = 0, \\
  \pdif{f}{m\subc^\mu} &= -J_0km\subc^\mu -\frac{i}{\beta}\hat{m}\subc^\mu = 0, \\
  \pdif{f}{m\subs^\mu} &= -J_0km\subs^\mu -\frac{i}{\beta}\hat{m}\subs^\mu = 0, \label{eq:saddle_ms}\\
  \pdif{f}{\hat{m}} &= -\frac{i}{\beta}\left(m - \frac{1}{2\pi}\int_{-\pi}^{\pi}\diff\theta \tanh\beta\tilde{h}(\theta) \right) = 0, \label{eq:saddle_hm}\\
  \pdif{f}{\hat{m}_0^\mu} &= -\frac{i}{\beta}\left(m_0^\mu - \frac{1}{2\pi}\int_{-\pi}^{\pi}\diff\theta \>\xi_{\theta}^\mu\tanh\beta\tilde{h}(\theta) \right) = 0, \\
  \pdif{f}{\hat{m}\subc^\mu} &= -\frac{i}{\beta}\left(m\subc^\mu - \frac{1}{2\pi}\int_{-\pi}^{\pi}\diff\theta \> \xi_{\theta}^\mu \cos\theta \tanh\beta\tilde{h}(\theta) \right) = 0, \\
  \pdif{f}{\hat{m}\subs^\mu} &= -\frac{i}{\beta}\left(m\subs^\mu - \frac{1}{2\pi}\int_{-\pi}^{\pi}\diff\theta \> \xi_{\theta}^\mu \sin\theta \tanh\beta\tilde{h}(\theta) \right) = 0. \label{eq:saddle_hms}
\end{align}
We can eliminate $\hat{m},\hat{m}_0^\mu,\hat{m}\subc^\mu,\hat{m}\subc^\mu$ from Eq. (\ref{eq:fe_before}) and Eqs. (\ref{eq:saddle_hm})--(\ref{eq:saddle_hms})
by using Eqs. (\ref{eq:saddle_m})--(\ref{eq:saddle_ms}). 
We then get 
\begin{equation}
  f = \frac{J_0}{2} \sum_{\mu=1}^p \left\{(m_0^\mu)^2 + k\bigl((m\subc^\mu)^2+(m\subs^\mu)^2\bigl) \right\} - \frac{g}{2}m^2
      - \frac{1}{2\pi\beta}\int_{-\pi}^\pi \diff\theta\log 2\cosh \beta\tilde{h}(\theta) ,
\end{equation}
\begin{align}
   m     &= \frac{1}{2\pi}\int_{-\pi}^{\pi} \diff \theta \tanh \beta \tilde{h}(\theta) \label{eq:sc_m}, \\
   m_0^\mu   &= \frac{1}{2\pi}\int_{-\pi}^{\pi} \diff \theta \>\xi_{\theta}^\mu \tanh \beta \tilde{h}(\theta) \label{eq:sc_m0}, \\
   m\subc^\mu &= \frac{1}{2\pi}\int_{-\pi}^{\pi} \diff \theta \> \xi_{\theta}^\mu \cos\theta \tanh \beta \tilde{h}(\theta) \label{eq:sc_mc}, \\
   m\subs^\mu &= \frac{1}{2\pi}\int_{-\pi}^{\pi} \diff \theta \> \xi_{\theta}^\mu \sin\theta \tanh \beta \tilde{h}(\theta) \label{eq:sc_ms}, 
\end{align}
where 
\begin{eqnarray}
  \tilde{h}(\theta)&=&J_0\sum_{\mu=1}^p\left(m_0^\mu+k(m\subc^\mu\cos\theta+m\subs^\mu\sin\theta) \right)\xi_\theta^\mu - g m + h \nonumber \\
                   &=&J_0\sum_{\mu=1}^p\left(m_0^\mu+km_1^\mu \cos(\theta-\phi^\mu)\right)\xi_\theta^\mu - g m + h.  \label{eq:h}
\end{eqnarray}

Next, we calculate the Hessian matrix at the saddle point. 
From the saddle point equations of conjugate variables in Eqs. (\ref{eq:saddle_hm})--(\ref{eq:saddle_hms}), 
we get $m,\{m_0^\mu\},\{m\subc^\mu\},\{m\subs^\mu\}$ as a function of 
$\hat{m},\{\hat{m}_0^\mu\},\{\hat{m}\subc^\mu\},\{\hat{m}\subs^\mu\}$: 
\begin{align}
   m     &= \frac{1}{2\pi}\int_{-\pi}^{\pi} \diff \theta \tanh \beta \tilde{h}(\theta), \label{eq:saddle_hm_tran} \\
   m_0^\mu   &= \frac{1}{2\pi}\int_{-\pi}^{\pi} \diff \theta \>\xi_{\theta}^\mu \tanh \beta \tilde{h}(\theta), \label{eq:saddle_hm0_tran} \\
   m\subc^\mu &= \frac{1}{2\pi}\int_{-\pi}^{\pi} \diff \theta \> \xi_{\theta}^\mu \cos\theta \tanh \beta \tilde{h}(\theta), \label{eq:saddle_hmc_tran} \\
   m\subs^\mu &= \frac{1}{2\pi}\int_{-\pi}^{\pi} \diff \theta \> \xi_{\theta}^\mu \sin\theta \tanh \beta \tilde{h}(\theta), \label{eq:saddle_hms_tran}
\end{align}
where $\tilde{h}(\theta)$ is given by Eq. (\ref{eq:h_before}). 
Using the inverse functions of Eqs. (\ref{eq:saddle_hm_tran})--(\ref{eq:saddle_hms_tran}), 
we can reduce the integrals in Eq. (\ref{eq:ZafterTr}):
\begin{equation}
    \int_{-\infty}^\infty \cdots\int_{-\infty}^\infty \diff \bm{m} \diff \hat{\bm{m}}
    \exp\left(-\beta N f(\bm{m}, \hat{\bm{m}})\right)
    \to 
    \int_{-\infty}^\infty \cdots\int_{-\infty}^\infty \diff \bm{m} 
    \exp\left(-\beta N f(\bm{m}, \hat{\bm{m}}(\bm{m}))\right), 
\end{equation}
where $\bm{m}$ and $\hat{\bm{m}}$ are 
$\left\{m,\{m_0^\mu\},\{m\subc^\mu\},\{m\subs^\mu\}\right\}$ and 
$\left\{\hat{m},\{\hat{m}_0^\mu\},\{\hat{m}\subc^\mu\},\{\hat{m}\subs^\mu\}\right\}$, respectively. 
Then, we can consider the Hessian matrix $G$ of the free energy $f(\bm{m}, \hat{\bm{m}}(\bm{m}))$: 
\begin{equation}
  G = 
  \begin{pmatrix}
    \pdif{f}{m} & \trp{\bm{u}_1} & \trp{\bm{u}_2} & \cdots & \trp{\bm{u}_p} \\
    \bm{u}_1    & G_{11}         & G_{12}         & \cdots & G_{1p} \\
    \bm{u}_2    & G_{21}         & G_{22}         & \cdots & G_{2p} \\
    \vdots      & \vdots         & \vdots         & \ddots & \vdots \\
    \bm{u}_p    & G_{p1}         & G_{p2}         & \cdots & G_{pp} \\
  \end{pmatrix}, 
\end{equation}
where 
\begin{align}
  \trp{\bm{u}_\mu} &= \begin{pmatrix} \pdifdif{f}{m}{m_0^\mu} &\pdifdif{f}{m}{m\subc^\mu} &\pdifdif{f}{m}{m\subs^\mu} \end{pmatrix}, \\
  G_{\mu\nu} &= 
  \begin{pmatrix}
    \pdifdif{f}{m_0^\mu}{m_0^\nu}    &\pdifdif{f}{m_0^\mu}{m\subc^\nu}    &\pdifdif{f}{m_0^\mu}{m\subs^\nu}    \\
    \pdifdif{f}{m\subc^\mu}{m_0^\nu} &\pdifdif{f}{m\subc^\mu}{m\subc^\nu} &\pdifdif{f}{m\subc^\mu}{m\subs^\nu} \\
    \pdifdif{f}{m\subs^\mu}{m_0^\nu} &\pdifdif{f}{m\subs^\mu}{m\subc^\nu} &\pdifdif{f}{m\subs^\mu}{m\subs^\nu} 
  \end{pmatrix}. 
\end{align}
From Eq. (\ref{eq:fe_before}) and Eqs. (\ref{eq:saddle_hm_tran})--(\ref{eq:saddle_hms_tran}), the Hessian matrix $G$ becomes 
\begin{equation}
  G = - J - \frac{i}{\beta}K^{-1}, 
\end{equation}
where
\begin{equation}
  J = 
  \begin{pmatrix}
    -g     & 0         & \cdots & 0 \\
    0      & \tilde{J} & \cdots & 0 \\
    \vdots & \vdots    & \ddots & \vdots \\
    0      & 0         & \cdots & \tilde{J}
  \end{pmatrix}, \>
  \tilde{J} = 
  \begin{pmatrix}
    J_0 & 0    & 0 \\
    0   & J_0k & 0  \\
    0   & 0    & J_0k
  \end{pmatrix}, \>
\end{equation}
and 
\begin{equation}
  K = 
  \begin{pmatrix}
    \pdif{m}{\hat{m}} & \trp{\bm{v}_1} & \trp{\bm{v}_2} & \cdots & \trp{\bm{v}_p} \\
    \bm{w}_1          & K_{11}         & K_{12}         & \cdots & K_{1p} \\
    \bm{w}_2          & K_{21}         & K_{22}         & \cdots & K_{2p} \\
    \vdots            & \vdots         & \vdots         & \ddots & \vdots \\
    \bm{w}_p          & K_{p1}         & K_{p2}         & \cdots & K_{pp}
  \end{pmatrix}, 
\end{equation}
where 
\begin{align}
    \trp{\bm{v}_\mu} &= \begin{pmatrix} \pdif{m}{\hat{m}_0^\mu} & \pdif{m}{\hat{m}\subc^\mu} & \pdif{m}{\hat{m}\subs^\mu} \end{pmatrix}, \\
    \trp{\bm{w}_\mu} &= \begin{pmatrix} \pdif{m_0^\mu}{\hat{m}} & \pdif{m\subc^\mu}{\hat{m}} & \pdif{m\subs^\mu}{\hat{m}} \end{pmatrix}, \\
    K_{\mu\nu} &= 
    \begin{pmatrix}
      \pdif{m_0^\mu}{\hat{m}_0^\nu}      & \pdif{m_0^\mu}{\hat{m}\subc^\nu}    & \pdif{m_0^\mu}{\hat{m}\subs^\nu} \\
      \pdif{m\subc^\mu}{\hat{m}_0^\nu}   & \pdif{m\subc^\mu}{\hat{m}\subc^\nu} & \pdif{m\subc^\mu}{\hat{m}\subs^\nu} \\
      \pdif{m\subs^\mu}{\hat{m}_0^\nu}   & \pdif{m\subs^\mu}{\hat{m}\subc^\nu} & \pdif{m\subs^\mu}{\hat{m}\subs^\nu}
    \end{pmatrix}. 
\end{align}
Let us consider the case where only the first stored pattern is retrieved. 
Since the order parameters of subsequent stored patterns $m_0^\mu, m\subc^\mu, m\subs^\mu (\mu\geq 2)$ are zero, 
and the stored patterns are orthogonal to each other, 
$\trp{\bm{v}_\mu}, \bm{w}_\mu \,(\mu\geq 2)$ and $K_{\mu\nu} \,(\mu\neq\nu)$ become zero. 
Furthermore, the diagonal blocks become equal to each other ($K_{11}=K_{22}=\cdots=K_{pp}$). 
Then, we can reduce the Hessian matrix $G$ to a block diagonal matrix: 
\begin{equation}
  G = 
  \begin{pmatrix}
    G_1     & 0      & \cdots & 0 \\
    0       & G_2    & \cdots & 0 \\
    \vdots  & \vdots & \ddots & \vdots \\
    0       & 0      & \cdots & G_2
  \end{pmatrix}
  =
  -J - \frac{i}{\beta}
  \begin{pmatrix}
    K_1^{-1}     & 0      & \cdots & 0 \\
    0       & K_2^{-1}    & \cdots & 0 \\
    \vdots  & \vdots & \ddots & \vdots \\
    0       & 0      & \cdots & K_2^{-1}
  \end{pmatrix}, \label{eq:G_last}
\end{equation}
where 
\begin{equation}
  K_1 = 
  \begin{pmatrix}
    \pdif{m}{\hat{m}} & \trp{\bm{v}_1} \\
    \bm{w}_1          & K_2
  \end{pmatrix}, 
\end{equation}
and $K_2=K_{11}=K_{22}=\cdots=K_{pp}$. 

\subsection{Mean-Field Equation}
In this section, we introduce the mean-field equation for later use. 
Since our model has infinite range interactions, the mean-field approximation is satisfied exactly within the thermodynamic limit. 
The mean-field equation is 
\begin{equation}
 \braket{S_{\theta_i}} = \tanh \beta \tilde{h}(\theta_i), \label{eq:mean-field}
\end{equation}
where $\tilde{h}(\theta)$ is given by Eq. (\ref{eq:h}). 
We can actually see the spin configurations from this equation in the following way. 
First, we calculate the values of the order parameters by solving 
Eqs. (\ref{eq:sc_m})--(\ref{eq:sc_ms}), and then 
substitute these values into the right-hand side of Eq. (\ref{eq:mean-field}). 
In this way, we will see the spin configurations presented in Section. {\ref{ch:Results}}. 

\section{Results}\label{ch:Results}
\begin{figure}
\includegraphics[width=6.4in]{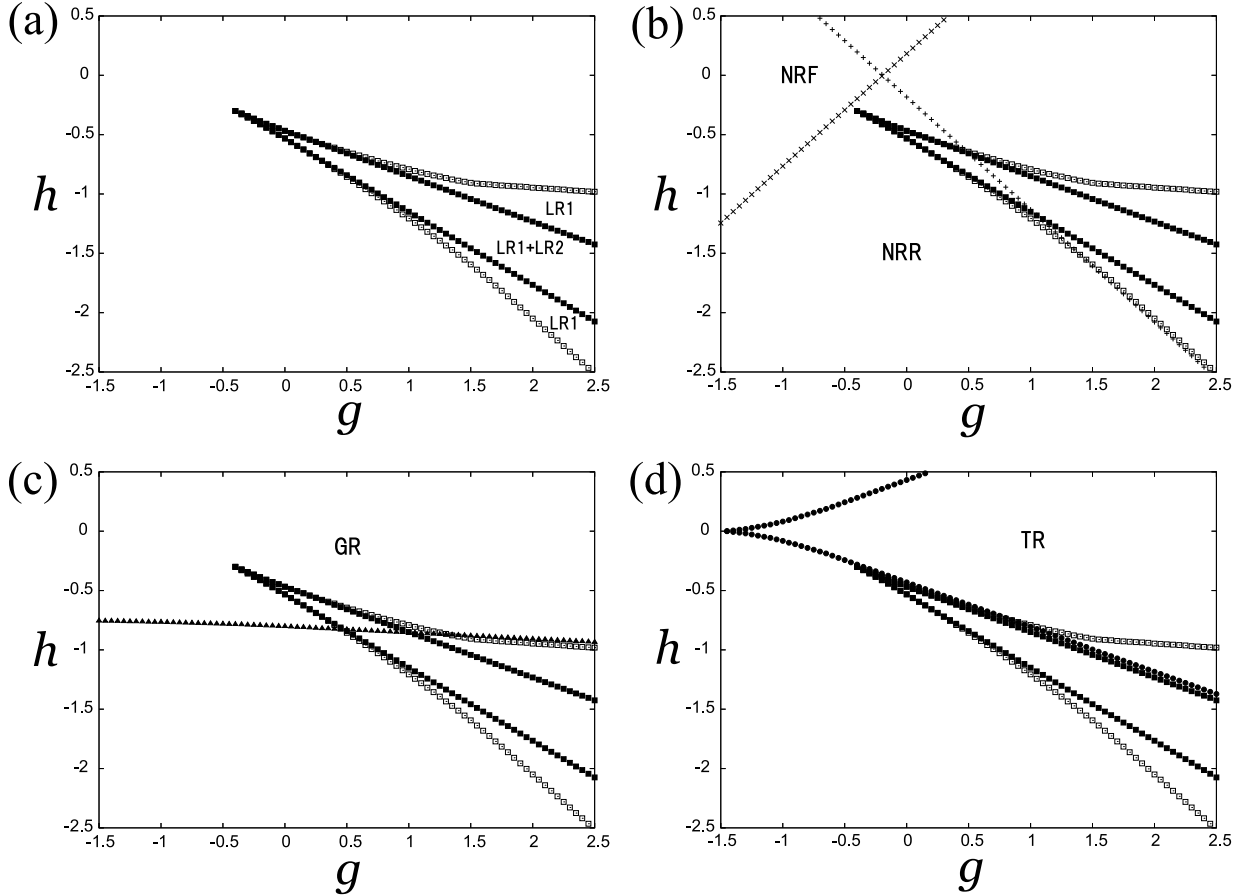}
\caption{
Phase diagram in $(g,h)$ space with fixed parameters $(\beta,J_0,k)=(10,1,1.5)$. 
There are five phases, 
the localized retrieval (LR) phase, 
non-retrieval-firing (NRF) phase, non-retrieval-resting (NRR) phase, 
global retrieval (GR) phase, and the twisted retrieval (TR) phase. 
(a) LR phase boundaries. 
The size of the LR phase depends on the number of stored patterns $p$. 
If $p=1$, the region of the LR phase is LR1. 
On the other hand, if $p\geq 2$, the region of the LR phase is LR2. 
The LR1 region covers LR2 region. 
In both cases, the range of $h$ of the LR phase expands when $g$ increases. 
The boundary of the LR1 phase and that of the LR2 phase are plotted 
using white squares ($\square$) and black squares ($\blacksquare$), respectively. 
(b)--(d) Positional relationships between LR phase and other phases. 
Crosses ($\times$) are used to plot the boundary of the NRF phase, 
pluses ($+$) for the NRR phase, 
triangles ($\blacktriangle$) for the GR phase, and 
bullets ($\bullet$) for the TR phase. 
}
\label{fig:phase}
\end{figure}

\begin{figure}
\includegraphics[width=3.2in]{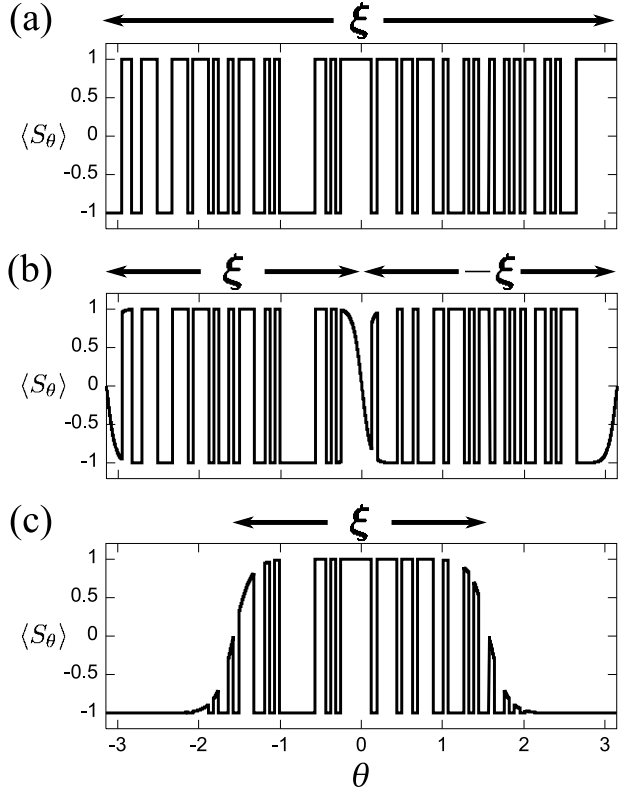}
\caption{
(a) Spin configuration in GR phase. 
A stored pattern is retrieved in the entire network; 
(b) Spin configuration in TR phase. 
One half of the spin configuration is aligned with the stored pattern, and 
the other half is a reverse to the stored pattern; and 
(c) Spin configuration in LR phase. 
A localized part of a stored pattern is retrieved in one half of the network, 
and spins are resting in the other half. 
}
\label{fig:GR_TR_LR_2}
\end{figure}

\begin{figure}
\includegraphics[width=3.2in]{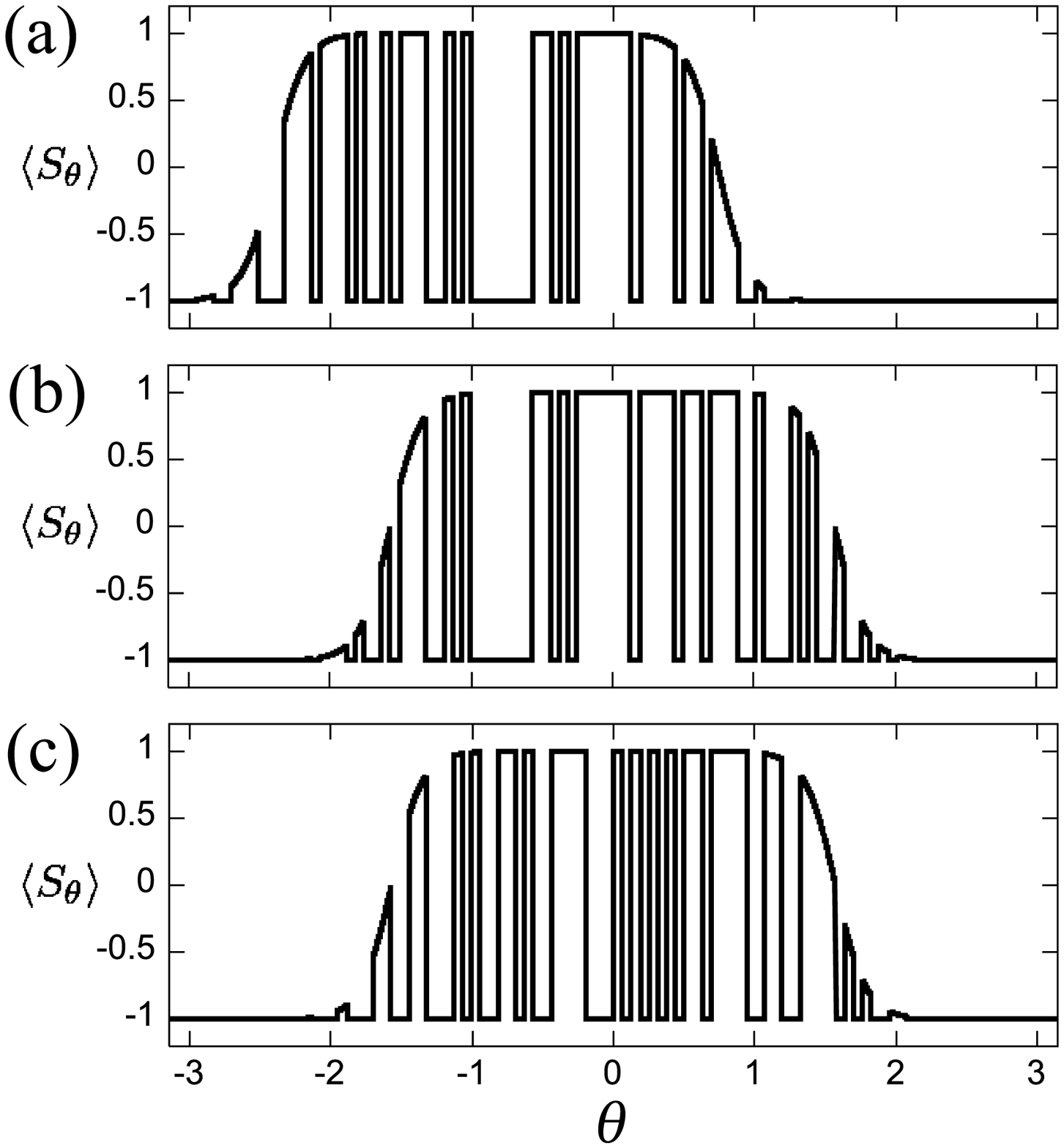}
\caption{
Spin configuration in LR phase with fixed parameters $(\beta,J_0,k,g,h)=(10,1,1.5,2,-1.5)$. 
(a) $\mu$th stored pattern is locally retrieved around location $\theta=-\pi/4$. 
(b) $\mu$th stored pattern is locally retrieved around location $\theta=0$. 
(c) $\nu(\neq \mu)$th stored pattern is locally retrieved around location $\theta=0$. 
}
\label{fig:neutral}
\end{figure}

\subsection{Phase Diagram}
In this subsection, 
we present the results from a bifurcation analysis with $(\beta,J_0,k)$ set to $(10,1,1.5)$ 
using Eqs. (\ref{eq:sc_m})--(\ref{eq:sc_ms}) and the Hessian matrix Eq. (\ref{eq:G_last}). 
For simplicity we confine ourselves to cases where only one or no stored pattern is retrieved. 
We derive five phases in our model. 
In two of the five phases, the stored patterns are not retrieved, and 
in the other three phases, a stored pattern is retrieved. 
In one of these three phases, our model achieves a discrete and continuous information representation.  

First, we give a detailed account of the two phases where stored patterns are not retrieved. 
In these two phases, the antiferromagnetic interaction $g$ and 
the external magnetic field $h$ govern the spin configuration. 
We distinguish between these two phases by looking at the sign of the magnetization $m$: 
the non-retrieval-firing (NRF) phase has a positive magnetization $m>0$ and 
the non-retrieval-resting (NRR) phase has a negative magnetization $m<0$. 
In the NRF phase, the firing neurons dominate the network, 
but the resting neurons dominate the network in the NRR phase. 
The NRF and NRR phase boundaries are shown in Fig. \ref{fig:phase}(b) 
represented as crosses ($\times$) and pluses ($+$), respectively. 

Second, we give a detailed account of the three phases where a stored pattern is retrieved.  
We label these three phases the global retrieval (GR), twisted retrieval (TR), and 
localized retrieval (LR) phases, which is 
the desired phase where our model achieves a discrete and continuous information representation. 
In the GR phase, 
a stored pattern is retrieved in the entire network and 
the spin configuration is not modulated by the Mexican-hat type interaction ($m_0^\mu>0, m_1^\mu=0, m_0^\nu=m_1^\nu=0\, (\nu\neq\mu)$) 
as shown in Fig. \ref{fig:GR_TR_LR_2}(a). 
The GR phase is equivalent to the retrieval phase of the Hopfield model, but 
we use the word "global" to distinguish the GR phase from the other two retrieval phases. 
Triangles ($\blacktriangle$) are used to represent the GR phase boundary in Fig. \ref{fig:phase}(c). 
In the TR phase, 
half of the spin configuration is aligned with a stored pattern, and 
the other half is a reverse of the stored pattern ($m_0^\mu=0,m_1^\mu>0, m_0^\nu=m_1^\nu=0\, (\nu\neq\mu)$), 
as shown in Fig. \ref{fig:GR_TR_LR_2}(b). 
Bullets ($\bullet$) are used to represent 
the TR phase boundary in Fig. \ref{fig:phase}(d). 
In the LR phase, 
a localized part of a stored pattern is retrieved in half of the network, 
and spins are resting in the other half ($m<0,m_0^\mu>0,m_1^\mu>0, m_0^\nu=m_1^\nu=0(\nu\neq\mu)$), 
as shown in Fig. \ref{fig:GR_TR_LR_2}(c). 
In the LR phase, $\phi^\mu$ represents the position of the localized retrieval. 
In both Fig. \ref{fig:neutral}(a) and Fig. \ref{fig:neutral}(b), 
a localized part of the $\mu$th stored pattern is retrieved, but 
the positions of the localized retrieval $\phi^\mu$ are different from each other. 
$\phi^\mu=-\pi/4$ in Fig. \ref{fig:neutral}(a), and $\phi^\mu=0$ in Fig. \ref{fig:neutral}(b). 
Thus, the LR states are neutrally stable along the positions of the localized retrieval. 
In Fig. \ref{fig:neutral}(b) and in Fig. \ref{fig:neutral}(c), 
the $\mu$th stored pattern and the $\nu$th stored pattern are locally retrieved around the same position, respectively. 
Thus, our proposed network can store and locally retrieve a number of patterns at the same position. 
The size of the LR phase depends on the number of stored patterns $p$. 
Let us consider the case where the first stored pattern is locally retrieved. 
We can confirm through an eigenvalue analysis that $G_1$ in Eq. (\ref{eq:G_last}) has 
a zero-eigenvalue corresponding to the neutral stability of the LR states on a one-dimensional ring 
(subsection \ref{ch:info}). 
The other eigenvalues of $G_1$ are positive in the entire LR1 phase. 
Therefore, if only one pattern is stored in the network ($p=1$), 
the LR states are stable throughout the entire LR1 phase. 
On the other hand, $G_2$ in Eq. (\ref{eq:G_last}) has a negative eigenvalue in part of the LR1 phase, 
that is, 
the difference between the LR1 and LR2 phases. 
Therefore, if there is more than one stored pattern ($p\geq2$), 
the LR states are stable only in the LR2 phase. 
Thus, we get the phase diagram shown in Fig. \ref{fig:phase}(a). 
If $p=1$, the LR phase region is LR1. 
On the other hand, if $p\geq 2$, the LR phase region is LR2. 
We must note that the LR1 region covers the LR2 region. 
In both cases, the range of $h$ of the LR phase expands as $g$ increases. 
This means that the antiferromagnetic interaction $g$ increases the robustness. 
We dwell on the features of the LR state in subsection \ref{ch:info}. 
The LR states represent the discrete and continuous information. 
\subsection{Role of Antiferromagnetic Interaction and External Magnetic Field}
\begin{figure}
\includegraphics[width=3.2in]{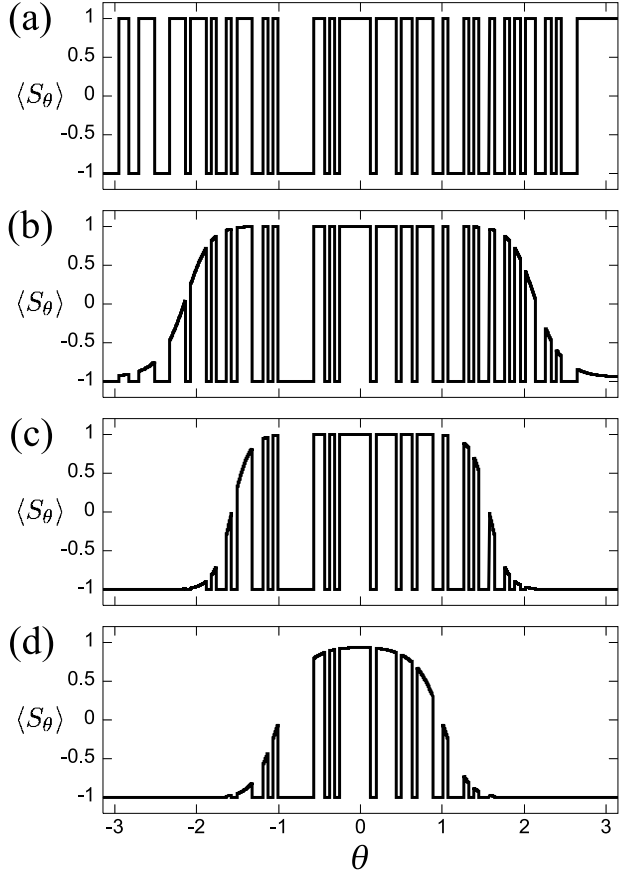}
\caption{
Spin configuration with fixed parameters $(\beta,J_0,k,g)=(10,1,1.5,2)$. 
(a) Spin configuration in GR phase with $h=-0.7$. 
(b)--(d) Spin configurations in LR phase. 
$h$ is set to $-1.1$, $-1.5$, and $-1.9$ in (b) to (d), respectively. 
With a decrease in $h$, the retrieval is localized to a narrower region. 
}
\label{fig:increase-h}
\end{figure}
In this subsection, we confirm the role of the external magnetic field $h$ and 
the antiferromagnetic interaction $g$ in the LR phase by using a mean-field analysis. 
$\tilde{h}(\theta_i)$ in the mean-field equation (\ref{eq:mean-field}) represents the local magnetic field 
applied to the spin located at $\theta_i$. 
We set $g>0$, and $h<0$. 
Let us consider the case where the $\mu$th stored pattern is locally retrieved. 
Then we have $m<0$, $m_0^\mu>0$, $m_1^\mu>0$, and $m_0^\nu=m_1^\nu=0\,(\nu\neq\mu)$. 
The local magnetic field becomes
\begin{equation}
 \begin{split}
  \tilde{h}(\theta_i)&=J_0\left(m_0^\mu+km_1^\mu \cos(\theta_i-\phi^\mu)\right)\xi_{\theta_i}^\mu - g m + h. \label{eq:effective}
 \end{split}
\end{equation}

First, we show the role of the external magnetic field $h$ in the LR phase. 
We focus on the term $J_0\left(m_0^\mu+km_1^\mu \cos(\theta_i-\phi^\mu)\right)\xi_{\theta_i}^\mu$, 
which is proportional to the $i$th component of the $\mu$th stored pattern vector $\bm{\xi}^\mu$. 
This term represents the magnetic field applied in the same direction as $\xi_{\theta_i}^\mu$ 
weighted by $J_0\left(m_0^\mu+km_1^\mu \cos(\theta_i-\phi^\mu)\right)$. 
Therefore, around the location $\theta_i=\phi^\mu$, $J_0\left(m_0^\mu+km_1^\mu \cos(\theta_i-\phi^\mu)\right)\xi_{\theta_i}^\mu$ 
attempts to strongly align the spin configuration with the $\mu$th stored pattern, 
whereas it weakly attempts this around the location $\theta_i=\phi^\mu\pm\pi$. 
Consequently, in the portion where the condition $J_0\left(m_0^\mu+km_1^\mu \cos(\theta_i-\phi^\mu)\right)<gm-h$ is satisfied, 
the local magnetic field has a negative value independent of $\xi_\theta^\mu$, 
leading to $\braket{S_{\theta_i}}<0$. 
Therefore, a decrease in $h$ causes an increase in $gm-h$, localizing a retrieval (Fig. \ref{fig:increase-h}(a)--(d)). 
Thus, the external magnetic field $h$ is essential for the LR state to exist. 

Second, we focus on the term $gm$ in Eq. (\ref{eq:effective}), and show the role of $g$ in the LR phase. 
We assume that a small displacement of the magnetization $m\to m+\Delta m$ is caused by noise. 
If we have a positive displacement $\Delta m>0$, the term $-gm$ decreases by $-g\Delta m$ to $-g(m+\Delta m)$. 
This means that a positive displacement of $m$ from the equilibrium 
additively generates a negative magnetic field, leading to displacement correction. 
In the same way, a negative displacement of $m$ from the equilibrium 
additively generates a positive magnetic field, leading to displacement correction. 
Thus, the antiferromagnetic interaction $g$ stabilizes the LR states. 

\subsection{Information Representation in LR Phase}\label{ch:info}
In this subsection, we discuss the information representation in the LR phase. 
Figure \ref{fig:neutral} shows three spin configuration instances in the LR phase. 

In Fig. \ref{fig:neutral}(a), the $\mu$th stored pattern is locally retrieved 
around the location $\theta=-\pi/4$, and 
in Fig. \ref{fig:neutral}(b), the $\mu$th stored pattern 
around the location $\theta=0$. 
There is a nonzero overlap between 
the spin configuration in Fig. \ref{fig:neutral}(a) and the one in Fig. \ref{fig:neutral}(b), 
because these two spin configurations coincide with each other in $-\pi/2<\theta<\pi/4$ region. 
Consequently, the information represented in Fig. \ref{fig:neutral}(a) and that in Fig. \ref{fig:neutral}(b) 
are associated with each other. 
Therefore, the localized retrievals of an individually stored pattern are distinguished and related to each other by the phase values $\phi^\mu$, 
that is, the localized retrievals form a ''group'', continuously ordered by the phase values $\phi^\mu$. 
Thus, a continuous information representation is achieved in the LR phase as well as in the model with the Mexican-hat type interaction. 

On the other hand, the $\nu$th stored pattern is locally retrieved around the location $\theta=0$ in Fig. \ref{fig:neutral}(c). 
Therefore, the spin configuration in Fig. \ref{fig:neutral}(b) and that in Fig. \ref{fig:neutral}(c) 
have an equal phase value ($\phi^\mu=\phi^\nu=0$). 
Additionally, there is no overlap between the spin configuration in Fig. \ref{fig:neutral}(b) and that in Fig. \ref{fig:neutral}(c), 
because of the orthogonality between $\bm{\xi}^\mu$ and $\bm{\xi}^\nu$. 
Consequently, the information represented in Fig. \ref{fig:neutral}(a) and that in Fig. \ref{fig:neutral}(b) 
are not associated with each other. 
Therefore, two localized retrievals from two different stored patterns have no overlap 
that is independent of the phase values. 
Thus, a discrete information representation is achieved in the LR phase as well as in the Hopfield model. 

Finally, we will now discuss the behavior of our model in the LR phase from the perspective of an attractor network. 
Let us consider the case where the $\mu$th stored pattern is retrieved 
($m<0$, $m_0^\mu>0$, $m_1^\mu>0$, and $m_0^\nu=m_1^\nu=0\,(\nu\neq\mu)$). 
The free energy is
\begin{equation}
  f = \frac{J_0}{2} \left\{(m_0^\mu)^2 + k(m_1^\mu)^2 \right\} - \frac{g}{2}m^2
      - \frac{1}{2\pi\beta}\int_{-\pi}^\pi \diff\theta\log 2\cosh \beta\tilde{h}(\theta) ,
 \label{eq:f_gl}
\end{equation}
where 
\begin{equation}
 \tilde{h}(\theta)=J_0\left(m_0^\mu+km_1^\mu \cos(\theta-\phi^\mu)\right)\xi_\theta^\mu - g m + h.
\end{equation}
We find using this equation that the free energy is independent of the phase value $\phi^\mu$, that is, 
the LR states are neutrally stable on a one-dimensional ring, which is termed a line attractor. 
Therefore, our model represents continuous information with the LR states. 
On the other hand, because different patterns are orthogonal to each other, 
the respective line attractors of the stored patterns are discretely distributed in the network state space. 
Thus, our model also represents discrete information. 
As shown above, our model achieves a discrete and continuous information representation. 

\section{Conclusion}
We have presented the results from our study on an Ising spin neural network 
that has a Hebbian interaction weighted by a Mexican-hat type interaction, 
an antiferromagnetic interaction, and an external magnetic field. 
We obtained five phases, 
the non-retrieval-firing (NRF) phase, the non-retrieval-resting (NRR) phase, 
the global retrieval (GR) phase, the twisted retrieval (TR) phase, and the localized retrieval (LR) phase 
by performing a statistical-mechanical analysis. 
In the LR phase, a localized part of a stored pattern was retrieved. 
The LR states are neutrally stable on a line attractor, which corresponds to a continuous information representation. 
In addition, because the stored patterns are orthogonal to each other, 
the respective line attractors of the stored patterns are discretely distributed in the network state space, 
corresponding to a discrete information representation. 
Thus, our model achieves a discrete and continuous information representation in the LR phase. 

From the phase diagram, we found 
that the stability of the LR states depends on the number of stored patterns, and 
that not only a Hebbian interaction weighted by a Mexican-hat type interaction, 
but also an antiferromagnetic interaction and a negative external magnetic field are essential to 
the existence of the LR phase. 
Additionally, by performing a mean-field analysis, 
we found that the antiferromagnetic interaction and the negative external magnetic field 
stabilize the LR state. 
Therefore, the antiferromagnetic interaction and the negative external magnetic field 
are essential for achieving a discrete and continuous information representation.

\end{document}